# Verification of a binary fluid solidification model in the finite-volume flow solver

T. Wacławczyk[a], M. Schäfer[a]

[a]*Institute of Numerical Methods in Mechanical Engineering,*
*Technische Universität Darmstadt, Dolivostr.15, Germany*

Corresponding author: dr.-ing. Tomasz Wacławczyk Institute of Numerical Methods in Mechanical Engineering, Dolivostr. 15, 64293 Darmstadt, Germany. E-mail: twacl@fnb.tu-darmstadt.de. Phone number: (+49) 6151 16-2525. Fax number: (+49) 6151 16-4479.

**Abstract.** The aim of this paper is to verify the new numerical implementation of a binary fluid, heat conduction dominated solidification model. First, we extend a semi-analytical solution to the heat diffusion equation, next, the range of its applicability is investigated. It was found that the linearization introduced to the heat diffusion equation negatively affects the ability to predict solidus and liquidus lines positions whenever the magnitude of latent heat of fusion exceeds a certain value. Next, a binary fluid solidification model is coupled with a flow solver, and is used in a numerical study of Al-4.1%Cu alloy solidification in a two-dimensional rectangular cavity. An accurate coupling between the solidification model and the flow solver is crucial for the correct forecast of solidification front positions and macrosegregation patterns.

## Introduction

The main difficulty in numerical simulations of casting with simultaneous solidification of metal alloys (e.g. Al-4.1%Cu, Sn-10%Pb) is the broad range of length and time scales that must be reconstructed. In order to avoid direct simulation of crystals growth, in engineering applications, problems in the modeling of the smallest length scales are resolved by the introduction of a volume averaging or mixture assumption. The volume averaging of flow and heat transfer governing equations give rise to the Darcy-Brinkman model for interstitial fluids (Bars and Worster, 2006), whereas the mixture theory allows equations for solid-liquid systems to be derived (Bennon and Incropera, 1987; Kurz and Fischer, 1980; Worster, 1986). In both approaches the phase change diagram and the macrosegregation model supply the physics of the phase transition and growth of the mushy zone at the microstructure level, where the smallest length scale is determined by details of the dendrites geometry. After the volume averaging is performed, the smallest time scales are related to the flow phenomena in a bulk fluid and porous (mushy zone) regions. Their accurate reconstruction is crucial for the prediction of a macrosegregation pattern, i.e. the nonuniform concentration of a solute. Since macrosegregation can lead to a variety of casting defects impairing the final quality of the product, its determination, monitoring and minimization is of primary interest to casting engineers.

The largest time scales at the other end are in the order of hours or days. They are inherited from the need to model heat conduction during the cooling of cast parts, e.g. a ship engine block. Because alloys have low heat diffusivity and because macrosegregation needs to be predicted simultaneously, simulations of the casting and solidification processes in industrial configurations are computationally complex and time-consuming. A multiscale approach, introduced by averaging of the governing equations and by modeling the sub-grid processes, allows for a reliable approximation of the model equations. However, their solution requires careful monitoring of the numerical procedure. This involves making certain that the mesh resolution and grid nodes distribution is appropriate, as well as a properly selecting the solver parameters, in particular, to ensure accurate solution of the energy conservation equation directly coupled with a solidification model.

In this paper, we verify a new numerical implementation of the binary fluid solidification model, where the phase-change process is driven by heat conduction or/and thermal and solutal buoyancy. In the first part, a semi-analytical solution to the energy transport equation is extended for the case of purely heat diffusive solidification taking into account, variable in the mushy zone, heat conductivity coefficient. We then show the conditions under which good agreement between our numerical model and the semi-analytical solutions is achieved. In the second part, convergence studies of the newly implemented model in the case of aluminum copper Al-4.1%Cu alloy solidification in a rectangular cavity are carried out.

## Physical and mathematical models

The binary fluid solidification model is based on the experimentally determined phase change diagram shown in Fig. 1. It describes the coexistence of a liquid, a solid and a mushy zone developed as a result of a flat solidification front instability caused by constitutional supercooling (Worster, 1986). In the mushy zone, a species $B$ with larger concentration solidifies, whereas the solute species $C$ is rejected to the liquid solution. The solute $C$ rejection increases its local concentration in the binary liquid and decreases the local solidification temperature (see shaded region depicted in Fig. 1). Such local variations in solidification temperature lead to a solidification front instability and growth of dendrites forest, i.e. the mushy zone.

The borders of phase change regions are defined in terms of solidus $T_{sol} = max(T_F + m_L C_M/k_P, T_E)$ and liquidus $T_{liq} = T_F + m_L C_M$ lines as functions of the mixture solute concentration $C_M = C_L f_L + C_S f_S$ and thus the local temperature $T$. The solid mass fraction $f_S = 1 - f_L$ can be determined from the lever rule

$$f_S = \frac{C_M - C_L}{C_S - C_L} = \frac{1}{1 - k_P} \frac{T - T_{liq}}{T - T_F} \tag{1}$$

or from the Scheil model

$$f_S = \left(\frac{C_M}{C_L}\right)^{\frac{1}{1-k_P}} = 1 - \left[k_P \frac{T_{sol} - T_F}{T_F - T}\right]^{\frac{1}{1-k_P}} \tag{2}$$

where it was assumed that $C_S = k_P C_L$ (Bennon and Incropera, 1987; Kurz and Fischer, 1980; Worster, 1986). For brevity, in the present paper we use the lever rule given by Eq. (1). However, all derivations and results presented here can be also obtained for the Scheil model. The coupling of the thermodynamical relations with a flow field is carried out using the mixture model (Bennon and Incropera, (1987); Wang and Beckermann, (1993)). Therein, it is assumed that the solid and liquid phases are uniformly distributed inside of the representative control volume. The size of the representative control volume is large enough to set properties of the medium inside effectively constant. The latter constraint or direct volume averaging of the phases allows to define the mass $f_k = m_k/m$ and the volume $g_k = V_k/V$ fractions, where $k = L, S$; in the present work, we set $\rho_S = \rho_L$ and thus $g_k = f_k$.

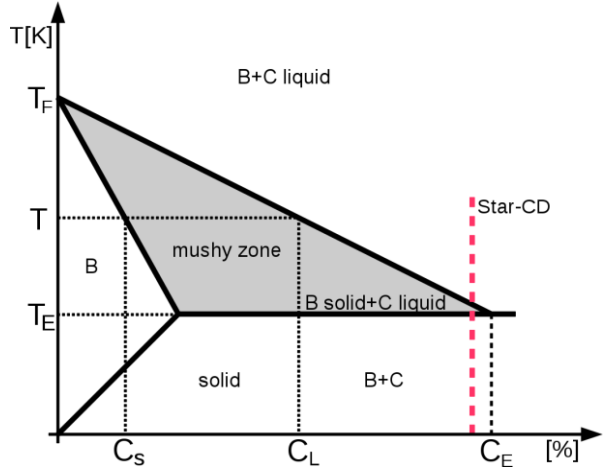

Fig. 1. A binary fluid phase change diagram, shaded region tags conditions (temperature and solute species $C$ concentration) for the mushy zone creation. $T_F$ denotes fusion temperature of the pure species $B$, $T_E$ and $C_E$ are eutectic temperature and solute concentration, $C_k$ where $k = L, S$ denote solute concentration in liquid and solid, respectively. The dashed, vertical line depicts schematically the default solidification model in the Star-CD flow solver $C = const$.

The thermophysical properties of the solid/liquid mixture are approximated with the arithmetic mean $\varphi = \varphi_S g_S + \varphi_L g_L$, where $\varphi$ represents any of the mixture material properties (density, specific heat, …) or mixture independent variables (velocity, enthalpy) in the conservation equations. The dynamic viscosity $\mu_L$ of the interstitial fluid in the mushy zone and the liquid alloy is assumed to be constant. Furthermore, the velocity of the solid $u_{i,s} = 0$ and therefore interaction between solid and interstitial fluid is restricted to the mushy zone. Here, Darcy's law with the Karman-Kozeny equation is used for modeling the pressure drop in the porous mushy region, see Eq. (3). When taking into account the aforementioned simplifications, the mixture continuity and the momentum equations become

$$\frac{\partial u_i}{\partial x_i} = 0 \qquad (3)$$

$$\frac{\partial \rho u_i}{\partial t} + \frac{\partial \rho u_i u_j}{\partial x_j} = \frac{\partial}{\partial x_j}\left[\mu_L\left(\frac{\partial u_i}{\partial x_j} + \frac{\partial u_j}{\partial x_i}\right)\right] - \frac{\partial p}{\partial x_i} - \frac{\mu_L}{K}u_i - \frac{\partial}{\partial x_j}\left[\rho(g_L u_{i,L} u_{j,L} - u_i u_j)\right]$$
$$+ \rho[1 - \chi(T - T_I) - \psi(C_L - C_I)]g_i \qquad (4)$$

where $u_j$ is the mixture velocity, $\chi$ and $\psi$ are thermal and solutal expansion coefficients, and the permeability $K$ is defined by the Karman-Kozeny equation $K = K_0 g_L^3/(1 - g_L)^2$ where $K_0 = \lambda^2/180$ is a material constant determined by the secondary dendrite arm spacing $\lambda$ (Samanta and Zabaras, 2004).

The energy equation governing heat transport during the solidification process can be written in terms of the mixture enthalpy

$$\frac{\partial \rho h}{\partial t} + \frac{\partial \rho u_j h}{\partial x_j} = \frac{\partial}{\partial x_j}\left(k \frac{\partial T}{\partial x_j}\right) - \frac{\partial}{\partial x_j}[\rho u_j (h_L - h)] \qquad (5)$$

defined by the arithmetic mean of the solid $h_S$ and liquid $h_L$ enthalpies $h = h_L g_L + h_S g_S$ where

$$h_S = c_S T \quad and \quad h_L = c_L T + L_{ht}. \qquad (6)$$

Finally, to reconstruct the solute concentration evolution in the mixture velocity field $u_i$ its transport equation is introduced. Since heat conduction dominated solidification is focus of the present work, we set $D_S = D_L = 0$ and therefore, the solute transport equation reads

$$\frac{\partial \rho C_M}{\partial t} + \frac{\partial \rho u_j C_M}{\partial x_j} = -\frac{\partial}{\partial x_j}[\rho u_j (C_L - C_M)]. \qquad (7)$$

A complete description and formal derivation of the binary fluid solidification model based on the mixture theory can be found in works of Bennon and Incropera (1987), Voller and Brent, (1989) Wang and Beckermann, (1993), whereas comprehensive derivation of the Darcy-Brinkman model is presented by Bars and Worster (2006). The model described above was implemented in the finite volume flow solver Star-CD with the aid of additional subroutines supplied by the code developer.

**A semi-analytical solution in a half-infinite domain.**

To verify the energy equation solver coupled with the binary fluid solidification model, we consider a semi-analytical solution for a one-dimensional heat diffusion dominated solidification in a half-infinite domain. In what follows, we shortly address some previous attempts to the semi-analytical solutions of the above problem and we present in details the approach introduced by Chakaraborty and Dutta (2002) which is extended in the present paper.

The semi-analytical solutions to the problem of unidirectional solidification of binary fluids were already investigated by Braga and Viscanta (1990), Chakaraborty and Dutta (2002) and Voller (1997) where a comprehensive overview of the semi-analytical solutions for the unidirectional solidification of binary fluids is given. In spite of this inherent drawbacks, Voller (1997) concludes the semi-analytical solutions can be considered as a useful tool for verification of numerical codes. The common feature of these semi-analytical solutions is the transformation of the partial differential energy conservation equation (5) into the set of ordinary differential equations using similarity variable $\eta = xg(t)$, $g(t) = 1/(2\sqrt{\alpha_S t})$. Since the resulting differential equations in the solid and liquid phase are linear, they are solved analytically therein. Obtained temperature profiles are used to specify heat fluxes at the solid and liquid sides of both solidification fronts, respectively. These fluxes are dependent on the non-dimensional positions of solidus $\eta_S$ and liquidus fronts $\eta_L$. The main difference between aforementioned works, lies in the approach to the solution of the heat diffusion equation in the mushy zone. Braga and Viscanta (1990) search only the position of $\eta_L$, therein, authors use data from the experiment to set the boundary conditions for temperature at the cold wall which is kept above eutectic temperature. After analytical solution of the heat diffusion equation in the liquid, the heat flux from the liquid side dependent on $\eta_L$ is used in the convergence criterion, i.e. the boundary condition for heat flux at liquidus line. The $\eta_L$ is obtained through corrections, in the iterative procedure, once the equality between fluxes on both front sides is achieved, $\eta_L$ is determined. We note that in work of Braga and Viscanta (1990) both $c_M$ and $k_M$ are variable inside the mushy zone what introduces the new non-linear terms in the heat diffusion equation. Voller (1997) uses another approach, in this work it was assumed $c_M = c_S = c_L$ and $k_M = k_S = k_L$ are constant in the whole domain, however, $\rho_S \neq \rho_L$ what results in a model taking in to account the velocity field generated by the shrinkage and macrosegregation. As Braga and Viscanta (1990), Voller (1997) also solves the heat diffusion equation numerically, avoiding linearization to obtain $\eta_S, \eta_L$. Chakaraborty and Dutta (2002) propose a different approach. Instead of the complex numerical procedure, the linearization of the heat diffusion equation in the mushy zone is introduced such that its analytical solution is possible. In what follows, we first present the semi-analytical solution introduced by Chakaraborty and Dutta (2002), and next, its modification is introduced allowing to take into account a variation of the heat conductivity coefficient in the mushy zone $k_M \neq k_S \neq k_L$.

Let us consider 1D solidification in the semi infinite domain $< 0, \infty)$, additionally, we assume that $\rho = const.$, $u_i = 0$, $c = const.$, $\partial k/\partial x \neq 0$. The enthalpies in solid $h_S$ and liquid $h_L$ are defined by Eqs. (6) and therefore, the mixture enthalpy reads

$$h = cT + (1 - g_S)L_{ht}. \tag{8}$$

Using the above assumptions and substituting Eq. (11) in to Eq. (5) we obtain

$$\rho c \frac{\partial T}{\partial t} - \rho L_{ht} \frac{\partial g_S}{\partial t} = \frac{\partial}{\partial x}\left(k \frac{\partial T}{\partial x}\right). \tag{9}$$

Because

$$\frac{\partial g_S}{\partial t} = \frac{\partial g_S}{\partial T}\frac{\partial T}{\partial t} \tag{10}$$

the problem of heat diffusion during solidification is described by the parabolic partial differential equation

$$\left(\frac{1}{\alpha_M} - \frac{\rho L_{ht}}{k_M}\frac{\partial g_S}{\partial T}\right)\frac{\partial T}{\partial t} = \frac{\partial^2 T}{\partial x^2} + \frac{1}{k_M}\frac{\partial k_M}{\partial x}\frac{\partial T}{\partial x} \tag{11}$$

where $g_S$ is calculated with Eq. (1) and $\alpha_M = k_M/\rho c_M$ is thermal diffusivity in the mushy zone. In the half-infinite domain, the boundary conditions for temperature are given in Tab. 1.

Table. 1. $T_C$ is temperature of the cold wall, $T_I$ is initial temperature, $x_S$ is the position of the solidus line and $x_L$ is the position of the liquidus line, $\theta_{liq} = (T_{liq} - T_{sol})/(T_F - T_{sol})$.

| x [m] | 0 | $x_S$ | $x_L$ | ∞ |
|---|---|---|---|---|
| T [K] | $T_C$ | $T_{sol}$ | $T_{liq}$ | $T_I$ |
| $\eta$ [−] | 0 | $\eta_S$ | $\eta_L$ | ∞ |
| $\theta$ [−] | $\theta_S = 0$ | $\theta_S = 1, \theta_M = 1$ | $\theta_M = \theta_{liq}, \theta_L = 1$ | $\theta_L = 0$ |

Additionally, conditions for jumps of heat fluxes at solid/mushy ($x = x_S$) and mushy/liquid ($x = x_L$) interfaces are

$$\text{at } x = x_S : \quad k_S \frac{\partial T_S}{\partial x} - k_M \frac{\partial T_M}{\partial x} = (1 - g_{S,sol})\rho L_{ht}\frac{dx_S}{dt} \tag{12}$$

$$\text{at } x = x_L : \quad k_M \frac{\partial T_M}{\partial x} - k_L \frac{\partial T_L}{\partial x} = g_{S,liq}\rho L_{ht}\frac{dx_L}{dt} \tag{13}$$

where $g_{S,sol}$ and $g_{S,liq}$ are solid fractions calculated for the temperatures $T_{sol}$ and $T_{liq}$ (see appendix C). To re-scale Eq. (11) nondimensional temperatures $0 \le \theta_k \le 1$ where $k = S, M, L$ are introduced, in a solid, a mushy zone and a liquid

$$\theta_S = \frac{T - T_C}{T_{sol} - T_C}, \quad \theta_M = \frac{T - T_{sol}}{T_F - T_{sol}}, \quad \theta_L = \frac{T - T_I}{T_{liq} - T_I}. \tag{14}$$

Next, the partial differential Eq. (11) is reduced to an ordinary differential equation, using a new similarity variable $\eta = xg(t)$, $g(t) = 1/(2\sqrt{\alpha_S t})$; the boundary conditions for non-dimensional temperatures $\theta$ are given in Tab. 1. Furthermore, the variable transformation (see appendix B) allows us to write Eq. (11) in the mushy zone ($\eta_S \le \eta \le \eta_L$) as

$$2\eta \frac{\alpha_S}{\alpha'_S} \frac{\partial \theta_M}{\partial \eta} + \frac{\partial^2 \theta_M}{\partial \eta^2} + \frac{1}{k_M} \frac{\partial k_M}{\partial \eta} \frac{\partial \theta_M}{\partial \eta} = 0 \tag{15}$$

where

$$\frac{\alpha_S}{\alpha'_S} = \frac{c_M - L_{ht} \frac{\partial g_S}{\partial T}}{c_S \frac{k_M}{k_S}}. \tag{16}$$

Since the volume fractions $g_S$ in the solid ($0 \leq \eta \leq \eta_S$) and $g_L = 1 - g_S$ in the liquid ($\eta_L \leq \eta \leq \infty$) are constant, we have $\partial g_S / \partial T = 0$ and $\alpha_S/\alpha'_S = 1$ in the solid phase and $\partial g_L/\partial T = 0$ and $\alpha_S/\alpha'_S = \alpha_S/\alpha_L$ in the liquid phase. Hence, the heat diffusion equations in the solid and liquid can be written as

$$2\eta \frac{\partial \theta_S}{\partial \eta} + \frac{\partial^2 \theta_S}{\partial \eta^2} = 0 \;:\; 0 \leq \eta \leq \eta_S \tag{17}$$

$$2\eta \frac{\alpha_S}{\alpha_L} \frac{\partial \theta_L}{\partial \eta} + \frac{\partial^2 \theta_L}{\partial \eta^2} = 0 \;:\; \eta_L \leq \eta \leq \infty \tag{18}$$

respectively. As Eqs. (17) and (18) are linear ODE's their direct analytical solution is possible. Because in the mushy zone $\partial g_S/\partial T$, $\alpha_M$, $k_M$ are dependent on the temperature, in order to obtain a linear form of Eq. (12), the averaging of the solid fraction derivative $\partial g_S/\partial T$ in Eq. (13)

$$\langle \frac{\partial g_S}{\partial T} \rangle = \frac{1}{\Delta T} \int_{T_{sol}}^{T_{liq}} \frac{\partial g_S}{\partial T} dT, \quad \Delta T = T_{liq} - T_{sol} \tag{19}$$

is proposed by Chakaraborty and Dutta (2002). Eq. (19) is later used for the calculation of the equivalent temperature $T_{eq}$ in the mushy zone from equation

$$\langle \frac{\partial g_S}{\partial T} \rangle = \frac{\partial g_S}{\partial T}. \tag{20}$$

When the solid fraction $g_S$ is calculated from the lever rule, see Eq. (1), the equivalent temperature reads

$$T_{eq} = T_F - \sqrt{(T_{liq} - T_F)(T_{sol} - T_F)} \tag{21}$$

if the Scheil model defined by Eq. (2) is used for $g_S$ calculation, one obtains

$$T_{eq} = T_F - [(1-k_P)A]^{-\frac{1-k_P}{2-k_P}}$$

$$A = \frac{1}{T_{liq} - T_{sol}} \left[ (T_F - T_{liq})^{-\frac{1}{1-k_P}} - (T_F - T_{sol})^{-\frac{1}{1-k_P}} \right]. \tag{22}$$

$T_{eq}$ calculated with Eq. (21) or Eq. (22) is now used to determine the equivalent solid fraction $g_{S,eq} = const$ and material properties $k_{M,eq} = const., c_{M,eq} = const$ representative for the whole mushy zone. One notices that when the heat conductivity $k_M = k_{M,eq}$ is constant, the assumption $\partial k_M/\partial \eta = 0$ used to derive Eq. (15) in is justified, see Chakaraborty and Dutta (2002). The averaging introduced in Eq. (19) and the employment of the boundary conditions allow for the analytical integration of Eqs. (15-18) with $\partial k_M/\partial \eta = 0$ in the mushy zone, in the solid and in the liquid separately. The result is temperature profiles given by

$$\theta_S = \frac{erf(\eta)}{erf(\eta_S)} \quad : \quad 0 \leq \eta \leq \eta_S \tag{23}$$

$$\theta_M = \theta_{liq} \frac{erf(\sqrt{a}\eta) - erf(\sqrt{a}\eta_S)}{erf(\sqrt{a}\eta_L) - erf(\sqrt{a}\eta_S)} \quad : \quad \eta_S \leq \eta \leq \eta_L \tag{24}$$

$$\theta_L = \frac{erfc(\sqrt{b}\eta)}{erfc(\sqrt{b}\eta_L)} \quad : \quad \eta_L \leq \eta \leq \infty \tag{25}$$

where b= $c_S k_L/c_L k_S$ and

$$a = \frac{\alpha_S}{\alpha'_S} = \frac{c_{M,eq} - L_{ht} \langle \frac{\partial g_S}{\partial T} \rangle}{c_S \frac{k_{M,eq}}{k_S}} \tag{26}$$

is the averaged coefficient in Eq. (15). Finally, the unknown positions of the solidus $\eta_S$ and liquidus $\eta_L$ lines are determined from the solution of the equation system obtained after substitution of Eqs. (23-25) into the boundary conditions for the jumps of heat fluxes

$$\theta_E \frac{\partial \theta_S}{\partial \eta} - r_{MS} \frac{\partial \theta_M}{\partial \eta} = \frac{2\eta_S(1 - g_{S,sol})}{St} \tag{27}$$

$$r_{MS} \frac{\partial \theta_M}{\partial \eta} - \theta_I r_{LS} \frac{\partial \theta_L}{\partial \eta} = \frac{2\eta_L g_{S,liq}}{St} \tag{28}$$

where $\theta_E = (T_{sol} - T_C)/(T_F - T_{sol})$, $r_{MS} = k_{M,eq}/k_S$, $\theta_I = (T_{liq} - T_I)/(T_F - T_{sol})$, $r_{LS} = k_L/k_S$, St= $c_S (T_F - T_{sol})/L_{ht}$ are given constants. In appendix C the derivation of jump conditions given by Eqs. (27-28) can be found.

## Modification of the semi-analytical solution

Since Eq. (5) is solved in the flow solver used in the present work, the influence of the term that contains the gradient of $k_M$ on the solidification process is always present. Therefore, a direct comparison between the semi-analytical solution introduced by Chakaraborty and Dutta (2002) and the numerical solution to Eq. (5) is not possible when $k_S \neq k_L$.

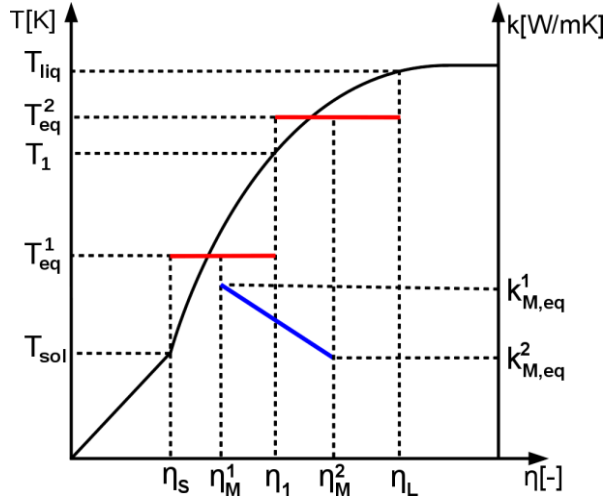

Fig. 2. A sketch of the modification to the semi-analytical solution, the modification is based on the introduction of the number of temperature intervals $N > 1$ for the averaging of the solid fraction temperature gradient in the mushy zone. In this figure $k_S > k_L$ and $N = 2$.

To generalize the solution of Eq. (15), we introduce a number of intervals $N > 1$ in which $T_{eq}^k$ is determined, $k = 1, \dots, N$; the complete description of the solution procedure for $N$ intervals can be found in appendix D. In the following, the case of two temperature intervals $N = 2$ is discussed as it is sufficient to present the main assumptions and problems of the modified semi-analytical solution. In the case of two intervals: $\langle T_{sol}, T_1 \rangle$, $\langle T_1, T_{liq} \rangle$ the averaging introduced by Eq. (19) can be carried out in each interval separately

$$\left\langle \frac{\partial g_S}{\partial T} \right\rangle \bigg|_1 = \frac{1}{\Delta T_1} \int_{T_{sol}}^{T_1} \frac{\partial g_S}{\partial T} dT, \quad \Delta T_1 = T_1 - T_{sol} \tag{29}$$

$$\left\langle \frac{\partial g_S}{\partial T} \right\rangle \bigg|_2 = \frac{1}{\Delta T_2} \int_{T_1}^{T_{liq}} \frac{\partial g_S}{\partial T} dT, \quad \Delta T_2 = T_{liq} - T_1 \tag{30}$$

where $T_1$ is the unknown temperature inside of the mushy zone as shown in Fig. 2. Next Eq. (20) must be solved in the two temperature intervals, which allows the calculation of the equivalent temperatures

$$T_{eq}^1 = T_F - \sqrt{(T_{liq} - T_F)/A_1} \tag{31}$$

$$T_{eq}^2 = T_F - \sqrt{(T_{liq} - T_F)(T_1 - T_F)} \tag{32}$$

where

$$A_1 = \frac{1}{T_{liq} - T_1}\left[\frac{T_1 - T_{liq}}{T_1 - T_F} - \frac{T_{sol} - T_{liq}}{T_{sol} - T_F}\right], \tag{33}$$

see appendix C for details. Using the equivalent temperatures $T_{eq}^1$ and $T_{eq}^2$ constant in each interval, we calculate the solid volume fractions $g_{S,eq}^1$ and $g_{S,eq}^2$. and material properties for each temperature interval inside of the mushy zone, e.g. $k_{M,eq}^1$, $k_{M,eq}^2$ as is shown in Fig. 2. Here we note that now the $k_{M,eq}^k$ distribution inside the mushy zone is piecewise constant. Thus, the last term on the RHS of Eq. (15) can be approximated by forward or backward differences (central differences are used when $N > 3$ - see appendix D), so that

$$b_1 = \frac{1}{k_{M,eq}^1}\frac{\partial k_M^1}{\partial \eta}\bigg|_1 \approx \frac{1}{k_{M,eq}^1}\frac{k_{M,eq}^1 - k_{M,eq}^2}{\eta_M^1 - \eta_M^2} = \frac{k_S - k_L}{k_{M,eq}^1}\frac{g_{S,eq}^2 - g_{S,eq}^1}{\eta_M^2 - \eta_M^1}, \tag{34}$$

$$b_2 = \frac{1}{k_{M,eq}^2}\frac{\partial k_M^2}{\partial \eta}\bigg|_2 \approx \frac{1}{k_{M,eq}^2}\frac{k_{M,eq}^2 - k_{M,eq}^1}{\eta_M^2 - \eta_M^1} = \frac{k_S - k_L}{k_{M,eq}^2}\frac{g_{S,eq}^2 - g_{S,eq}^1}{\eta_M^2 - \eta_M^1}, \tag{35}$$

where $\eta_M^1 = (\eta_S + \eta_1)/2$, $\eta_M^2 = (\eta_1 + \eta_L)/2$. The above approximation corresponds to the assumption of a constant heat conductivity gradient $\partial k_M/\partial x = \partial k_M/\partial \eta \, g(t) = const$ and, thus, a constant distribution of $k_{M,eq}$ in each temperature interval allows the introduction of a piecewise linear variation of the heat conductivity coefficient $k_M$ in the mushy zone. Using the latter assumption, Eq. (15) can be rewritten as

$$(2\eta a_1 + b_1)\frac{\partial \theta_M^1}{\partial \eta} + \frac{\partial^2 \theta_M^1}{\partial \eta^2} = 0 \; : \; \eta_S \leq \eta \leq \eta_1, \tag{36}$$

$$(2\eta a_2 + b_2)\frac{\partial \theta_M^2}{\partial \eta} + \frac{\partial^2 \theta_M^2}{\partial \eta^2} = 0 \; : \; \eta_1 \leq \eta \leq \eta_L \tag{37}$$

where $a_k$, $b_k$ for $k = 1,2$ are constants determined from Eq. (26) and Eqs. (34-35), respectively. Analytical solutions to Eqs. (36-37) read

$$\theta_M^1 = \theta_{liq}^1 \frac{erf[d_1(\eta)] - erf[d_1(\eta_S)]}{erf[d_1(\eta_1)] - erf[d_1(\eta_S)]} \tag{38}$$

$$\theta_M^2 = \theta_{liq}^2 \frac{erf[d_2(\eta)] - erf[d_2(\eta_1)]}{erf[d_2(\eta_L)] - erf[d_2(\eta_1)]} \tag{39}$$

where

$$d_k(\eta) = \sqrt{a_k}\eta + \frac{b_k}{2\sqrt{a_k}}, \quad k = 1,2 \tag{40}$$

and $\theta_M^1 = (T - T_{sol})/(T_F - T_{sol})$, $\theta_{liq}^1 = (T_1 - T_{sol})/(T_F - T_{sol})$, $\theta_M^2 = (T - T_1)/(T_F - T_1)$, $\theta_{liq}^2 = (T_{liq} - T_1)/(T_F - T_1)$. Here we note that the second term in Eq. (40) is inherited from the finite difference approximation of the heat conductivity gradient.

The remaining problem is how to determine the four unknowns depicted in Figure 2: $\eta_S, \eta_L, \eta_1$ and temperature in the mushy zone $T_1$. An equation required to determine $T_1$ is obtained from the condition of temperature continuity inside the mushy zone (see Eq. (B5) in appendix B). Hence, the boundary conditions for the heat fluxes at $\eta_S$, $\eta_1$ and $\eta_L$ become

$$\theta_E \frac{\partial \theta_S}{\partial \eta} - r_{MS}^1 \frac{\partial \theta_M^1}{\partial \eta} = \frac{2\eta_S(1 - g_{S,sol})}{St_0}, \tag{41}$$

$$\theta_{sol}^1 \frac{\partial \theta_M^1}{\partial \eta} - \frac{\partial \theta_M^2}{\partial \eta} = 0, \tag{42}$$

$$r_{MS}^2 \frac{\partial \theta_M^2}{\partial \eta} - \theta_I r_{LS} \frac{\partial \theta_L}{\partial \eta} = \frac{2\eta_L g_{S,liq}}{St_1}, \tag{43}$$

where $r_{MS}^1 = k_{M,eq}^1/k_S$, $r_{MS}^2 = k_{M,eq}^2/k_S$, $\theta_{sol}^1 = (T_F - T_{sol})/(T_F - T_1)$, $St_0 = c_S (T_F - T_{sol})/L_{ht}$ and $St_1 = c_S (T_F - T_1)/L_{ht}$. The derivation of Eqs. (41-43) is given in appendix C. Additionally, we notice that since the set of Eqs. (41-43) is non-linear, $\eta_1$ may be replaced by a linear combination of $\eta_S$ and $\eta_L$. Hence, for the two temperature intervals $\eta_1 = (\eta_S + \eta_L)/2$ and thus all four unknowns can be determined.

As in the original solution, the analytically obtained temperature profiles given by Eqs. (23, 25, 38, 39) are used to build a non-linear set of equations after substitution into the boundary conditions given by Eqs. (41-43) (see appendix D for more details). The obtained set of equations is solved using the Matlab *fsolve* non-linear equation solver, which uses either the Trust-Region Dogleg or Newton-Rhapson method. We found that the results, i.e. values of $\eta_S$, $T_1$ and $\eta_L$ do not depend on the solver used to obtain the solution. When $N = 2$, the initial value of the temperature inside the mushy zone $T_1^I$ is set to $T_1^I = (T_{liq} + T_{sol})/2$. When $N \geq 3$ the initial values of the temperatures in the mushy zone sub-division points $T_l^I$ where $l = 2, ..., N - 1$ are determined from the temperature profiles obtained in the previous iteration. The determination of subsequent $T_l^I$ values is crucial for the convergence of the solution procedure with an increasing number of intervals.

The above semi-analytical solution is generalized with respect to the number of sub-domains $N \leq 8$ and its convergence is investigated. In Figure 3a, the decreasing distances between the positions of the solidus $\eta_S$ and liquidus lines $\eta_L$ obtained with successively increasing numbers of temperature intervals $N$ are depicted. Using solutions on successively refined grids, we are able to determine an interval number $N$ independent solution with the aid of Richardson extrapolation (Schäfer, 2006). A comparison with a numerical solution obtained using the Star-CD solver is shown in Figure 3b.

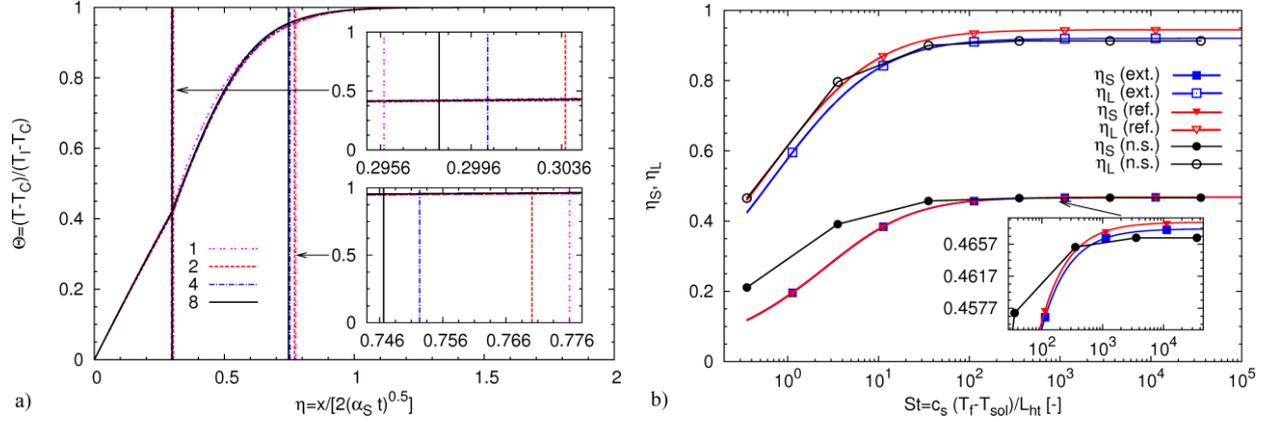

Fig. 3. a) Temperature profiles obtained with different numbers of averaging intervals $N = 1, 2, 4, 8$. b) A comparison of the semi-analytical solution with N=1 (ref.), modified and interval number $N$ independent semi-analytical solution (ext.) and numerical solution from the Star-CD code (n.s.).

**Verification of the numerical model compared to the semi-analytical solution**

The solidification algorithm was initially verified using material parameters from Voller and Brent (1989) for an aqueous solution of ammonium chloride $NH_4 70\% H_2O$ (see Tab. 2) and boundary conditions for temperature at $\eta = 0$, $T_C = 220\ K$ and $\eta \to \infty$, $T_C = 310\ K$ as in work of Chakaraborty and Dutta (2002).

Table 2. The material properties of an aqueous solution of ammonium chloride $NH_4 70\% H_2O$.

| $c[J/kgK]$ | $\rho[kg/m^3]$ | $k_S[W/m^2K]$ | $k_L[W/m^2K]$ | $L_{ht}\ [J/kg]$ | $T_F[K]$ | $T_{sol}$ | $T_{liq}$ |
|---|---|---|---|---|---|---|---|
| 3000 | 1078 | $T_{sol}$ | $T_{liq}$ | $3.18 \cdot 10^5$ | 633.59 | 257.95 | 305.95 |

The similarity solutions presented in Figs. 3-4 are determined at time $t = 5000\ s$. Because of the disagreement between semi-analytical and numerical solutions obtained for this set of thermophysical parameters, we studied the influence of the $L_{ht}$ magnitude on the positions of the solidus and liquidus lines. We emphasize that $L_{ht}$ is a continuous parameter in Eq. (26). This study allows us to find a range of Stefan numbers where the original semi-analytical solution ($k_S = k_L = k_M$) and its extension ($k_S \neq k_L \neq k_M$) can be used in the verification procedure. The range of the applicability of the semi-analytical solution was not discussed by Chakaraborty and Dutta (2002). In Figure 3b we observe that when $L_{ht} \leq 1.12 \cdot 10^4 J/kg$, which corresponds to a Stefan number $St \geq 100$, the agreement between the interval number independent, semi-analytical solution and the numerical solution is satisfactory. Moreover, the difference between $\eta_S$, $\eta_L$ positions predicted by the flow solver and the modified semi-analytical solution is smaller than the difference between the original solution and the present numerical results. This confirms the need to take into account the variation of the heat conductivity coefficient $k_M$ in the semi-analytical solution of Eq. (11).

When larger values of $L_{ht} \geq 1.12 \cdot 10^4 J/kg$ are used, i.e. $St \leq 100$ the positions of the solidus line $\eta_S$ predicted by the original or modified semi-analytical solutions and the numerical simulation disagree. The coincident shift of the solidus and liquidus lines towards the cold wall can be explained by the retardation of the solidification process caused by the rejection of a larger amount of the latent heat, see Figure 3b. However, in the case of semi-analytical solutions, e.g. for $L_{ht} = 3.18 \cdot 10^5 J/kg$, i.e. $St \approx 3.5$, the change of the solidus line position is too rapid. In the forthcoming section we will show how the removal of the non-linearity from Eq. (12) affects the $\eta_S$, $\eta_L$ positions.

**Influence of non-linearity in the heat diffusion equation on mushy zone thickness**

To assess the influence of the nonlinear term $\alpha_S/\alpha'_S$ in Eq. (15) on the shift of the solidus line for $L_{ht} \geq 1.12 \cdot 10^4 J/kg$, $St \leq 100$, we rewrite Eq. (15) as a set of two ordinary differential equations that can be solved as an initial value problem

$$\frac{\partial T}{\partial \eta} = \varphi, \tag{44}$$

$$\frac{\partial \varphi}{\partial \eta} = -\frac{2\eta \alpha_S}{\alpha'_S(T)} \varphi, \tag{45}$$

where the coefficient $\alpha_S/\alpha'_S$ is calculated directly with Eq. (16). The boundary conditions for the set of Eqs. (44-45) are given by $T = T_C$ for $\eta = 0$ and $T = T_I$ for $\eta \to \infty$. The latter boundary condition is obtained iteratively, because for the solution of the initial value problem both the temperature and the temperature gradient are required at $\eta = 0$. To remove effects associated with jumps of heat fluxes at the solidus and liquidus fronts (i.e. possible discontinuity in a temperature gradient profile, see Eqs. (12-13)) and disregard the influence of the term $\partial k/\partial x$ in this test case, we set the heat conductivity coefficient to the constant value $k_M = k_S = k_L = 0.4\ Wm^{-2}K^{-1}$ in the whole domain $0 \leq \eta \leq 8$.

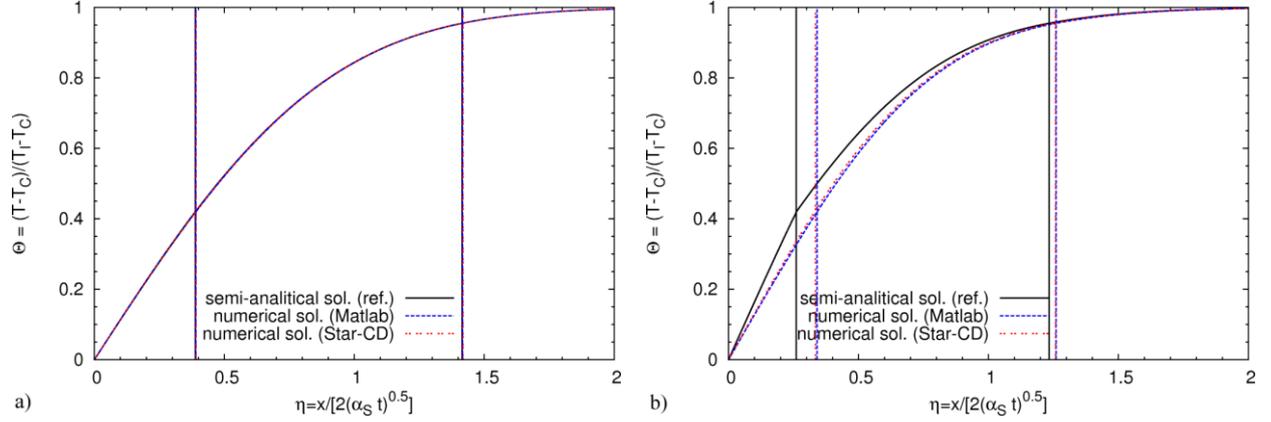

Fig. 4. Influence of the linearization procedure (temperature averaging) on the positions of solidus $\eta_S$ and liquidus $\eta_L$ lines. The comparison between reference semi-analytical solution $N = 1$ (solid line), numerical solution of the initial value problem from Matlab (dashed line), numerical solution of Eq. (5) from Star-CD (double-dotted line) obtained for two values of latent heat a) $L_{ht} = 3.18 \cdot 10^3 J/kg$, $St \approx 309$, b) $L_{ht} = 3.18 \cdot 10^5 J/kg$, $St \approx 3.5$.

The comparison of results obtained from the original semi-analytical solution $N = 1$, the numerical solutions of the initial value problem in Matlab and the solution of the general energy conservation equation (5) in Star-CD is depicted in Figure 4. In Figure 4a, where $L_{ht} = 3.18 \cdot 10^3 J/kg$ and $St \approx 309$, we can observe good agreement between both numerical solutions and the original semi-analytical solution. The main observation in Figure 4b is the disagreement of the semi-analytical solution and both numerical solutions when $L_{ht} = 3.18 \cdot 10^5 J/kg$ and $St \approx 3.5$. This latter result coincides with results presented in Figure 3b. The second observation to be made from Figure 4 is the good agreement between both numerical solutions independently on the $L_{ht}$ magnitude. These two observations confirm the influence of the linearization introduced to the semi-analytical solution of Eq. (15) on the artificial shift of the solidus line $\eta_S$ and thus the incorrect prediction of the mushy zone thickness and solidification fronts positions when $St \leq 100$.

**Solidification of Al-4.1%Cu alloy in rectangular cavity**

After the successful verification of the numerical solution to Eq. (5) compared with the semi-analytical solution given by Eqs. (D15-D17) and numerical solution of Eqs. (44-45), the flow model described by the set of conservation Eqs. (3-7) is used with the solidification model (see also Jakumait et al. (2012), Wacławczyk et al. (2011)). To verify our model implementation in the presence of thermosolutal buoyancy, solidification of Al-4.1%Cu alloy is modeled in a two-dimensional, rectangular cavity cooled from one side, see Figure 5 and Table 3.

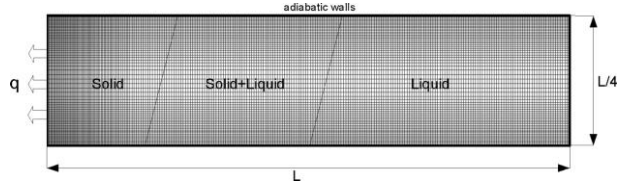

Fig. 5. The computational domain with $m_1: 64 \times 256$ grid and boundary conditions for Al-4.1%Cu alloy solidification, $L = 0.08\ m$, heat flux extracted from the left wall $q = h_{conv}(T - T_{amb})$, where $h_{conv} = 1000\ W/m^2 K$, $T_{amb} = 293.15\ K$, $T$ is temperature of the cooled wall.

Table 3. Material properties of Al-4.1%Cu from Samanta and Zabaras (2004).

| $C_I[\%]$ | $C_E[\%]$ | $k_P[-]$ | $\chi[1/K]$ | $\psi[1/K]$ | $K_0[m^2]$ | $L_{ht}[J/kg]$ | $T_E[K]$ | $T_F[K]$ |
|---|---|---|---|---|---|---|---|---|
| 4.1 | 33.21 | 0.173 | $4.95 \cdot 10^{-5}$ | $-2.0$ | $5.6 \cdot 10^{-11}$ | $3.97 \cdot 10^5$ | 821.1 | 933.6 |

| Phase | $c[J/kgK]$ | $k[W/m^2K]$ | $\rho[kg/m^3]$ | $\mu[kg/ms]$ |
|---|---|---|---|---|
| S | 1058.8 | 82.61 | 2525 | - |
| L | 1092.8 | 192.49 | 2525 | $3.0 \cdot 10^{-3}$ |

Because in the present study diffusivity of the solute is neglected in Eq. (7), the solutal Rayleigh number $Ra = \rho_L g_2 \psi \Delta C H^3/(\mu_L D_L)$ (where $H = L/4$ - see Figure 5) and solutal Schmidt number $Sc = \mu_L/(\rho_L D_L)$ can be considered as equal to infinity. The values of the thermal Rayleigh number $Ra_T = \rho_L g_2 \chi \Delta T H^3/(\mu_L \alpha_L)$ and the thermal Prandtl number Pr= $\mu_L/(\rho_L \alpha_L)$ can be estimated as $Ra_T \approx 1.27 \cdot 10^4$ and $Pr \approx 3.84 \cdot 10^{-2}$ taking the temperature difference $\Delta T = 133\ K$ from the end of the simulation (see Fig. 11, $t = 121\ s$). The values of nondimensional numbers locate the current simulation in a region typical for metal alloys simulations, i.e. $Ra_T \gg 1$ and $Pr \ll 1$.

Present numerical studies are performed on three gradually refined grids $m_1: 64 \times 256$, $m_2: 96 \times 384$, $m_3: 128 \times 512$ CV's, where the CFL condition is satisfied. To determine the influence of the solver settings on the evolution of macrosegregation, two monitoring functions are used. The global extent of macrosegregation

$$GES = \frac{1}{C_I}\left[\sum_{j=1}^{N_c}\frac{\left(C_{M,j}-C_I\right)^2}{N_c}\right]^{1/2} \tag{46}$$

is an integral measure which allows us to assess the extent of the variation of $C_M$ from its initial value $C_I$ in the computational domain discretized using a mesh with $N_c$ cells. The second monitoring function is the difference between the maximum and minimum mixture concentration

$$\Delta C = C_M^{max} - C_M^{min}. \tag{47}$$

The latter parameter indicates the local variation of the mixture concentration $C_M$ in regions maximally enriched and depleted by the solute. GES and $\Delta C$ are not independent since a larger variation of the concentrations difference also affects the value of the global extent of segregation GES (compare Eq. (46) and Eq. (47)).

Initially, we perform an evaluation of the solution sensitivity on the accuracy to which the energy equation (5) is solved. The accuracy of coupling between the energy equation and the solidification model is determined by the number of outer iterations per time step. The recognition of this issue is indispensable since in Star-CD both the mixture enthalpy $h$ and the temperature $T$ are used in the energy equation. For this reason a procedure allowing the recalculation of the temperature $T$ from the given enthalpy $h_g$ (which is the solution of Eq. (5) at the current time step and outer iteration) is required and is essential for the simulation process. In the aforementioned recalculation procedure we use a bi-section method to solve equation $h(T, g_S(T)) = h_g$ in each control volume. In Figure 6, profiles of the solid fraction $g_S$ and the mixture concentration $C_M$ after $t = 121\ s$ are shown, such that a variable parameter is the number of outer iterations per time step. These profiles were obtained from the simulation on the coarsest mesh $m_1$ and were stored in a horizontal cross section $y = 7.5\ mm$. In Figure 6 a strong dependence of the $g_S$ and $C_M$ profiles positions and shapes on the number of outer iterations per time step is observed. Since the $g_S$ profile is directly coupled with the temperature field (see Eq. (1)) and $C_M$ is coupled with velocity (see Eq. (7)), the convergence of these two quantities also confirms the convergence of temperature and velocity fields, respectively.

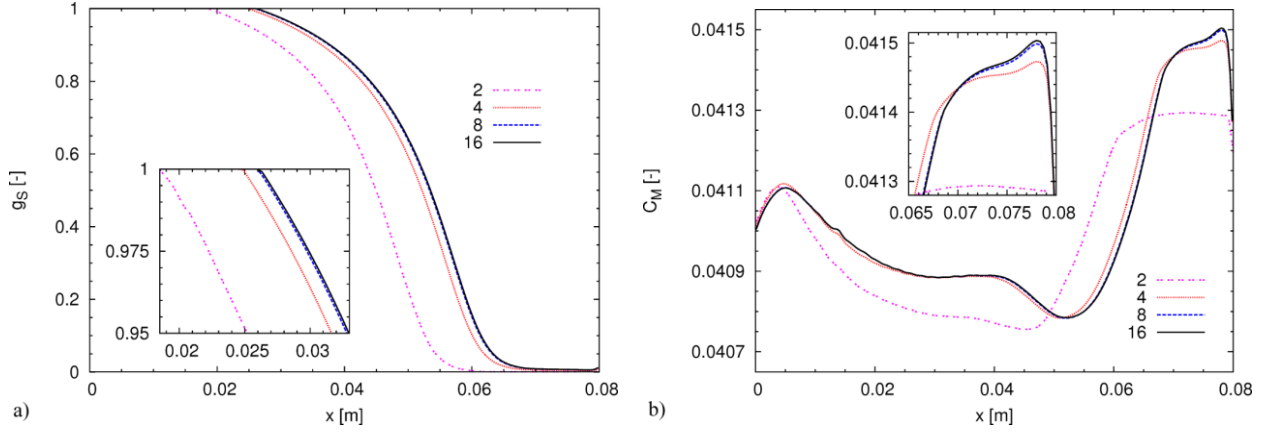

Figure 6. The convergence of a) the solid fraction $g_S$, b) the mixture concentration $C_M$. The profiles are obtained after $t = 121 s$ in a cross section $y = 7.5\ mm$ on $m_1: 64 \times 256$ grid, with increasing number of outer iteration $(2 − 16)$ per time step.

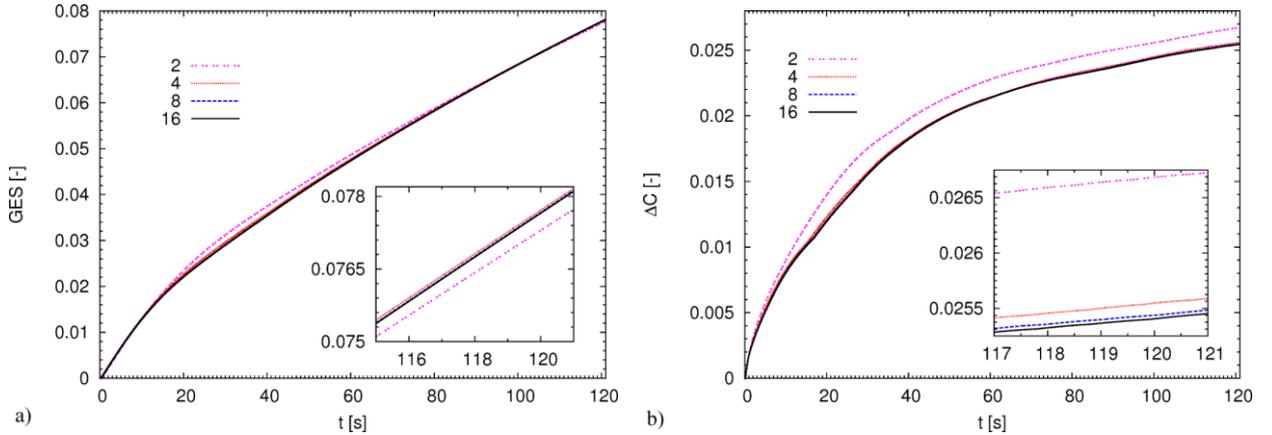

Figure 7. Histories of the convergence of a) the global extent of segregation GES, b) the concentrations difference $\Delta C$. The results are obtained for different numbers of outer iterations $(2 − 16)$ per time step, on $m_1: 64 \times 256$ CV's mesh.

The same conclusion regarding the minimal number of outer iterations per time step required for the convergence, can be drawn from Figure 7, where GES and $\Delta C$ histories of convergence are presented. Here, the convergence is also obtained for 8 or more outer iterations per time step. When the number of outer iterations is smaller, a decrement in GES and an increment in $\Delta C$ values is observed. The results depicted in Figure 6 and Figure 7 show that an accurate coupling between the solidification model and the flow solver is a crucial factor for the determination of reliable solidification front positions, mushy zone thickness and the solute distribution. All simulation results presented in the following sections are obtained with 8 outer iterations per time-step.

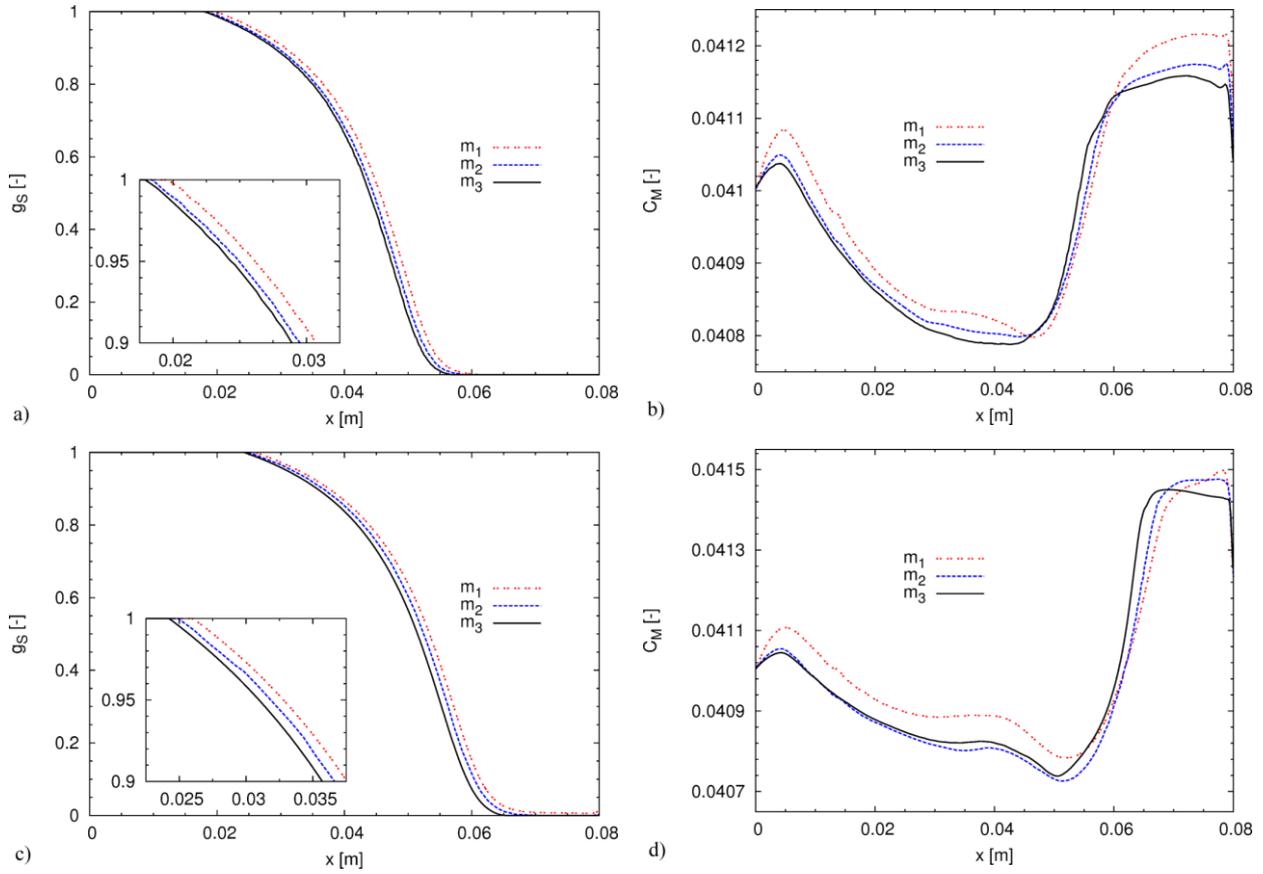

Figure 8. The convergence of a), c) the solid fraction $g_S$ and b), d) mixture concentration $C_M$ after $t = 100\ s$ a), b) and $t = 121\ s$ c), d) in a cross section $y = 7.5\ mm$, on three gradually refined grids $m_1: 64 \times 256$, $m_2: 196 \times 384$, $m_3: 128 \times 512$ CV's.

A comparison of the results obtained on three gradually refined grids is presented in Figures 8-9. To obtain a convergent solution it is necessary to refine the grid close to horizontal, adiabatic walls and the cold wall where a heat flux is extracted, see Figure 5. In Figure 8, convergence of the solid fraction $g_S$ and the mixture concentration $C_M$ profiles on three grids is depicted at two different times $t = 100\ s$ and $t = 121\ s$. At time moment $t = 100\ s$, see Figure 8a, decreasing distances between solid fraction profiles can be observed. At the final time moment $t = 121\ s$, the convergence of the solidification front positions is less evident, in particular near the liquidus solidification front, see Figure 8c. The reason for this is the presence of the right adiabatic wall, which affects heat transport and thus the shapes of the $g_S$ profiles obtained on the $m_1$ and $m_3$ grids.

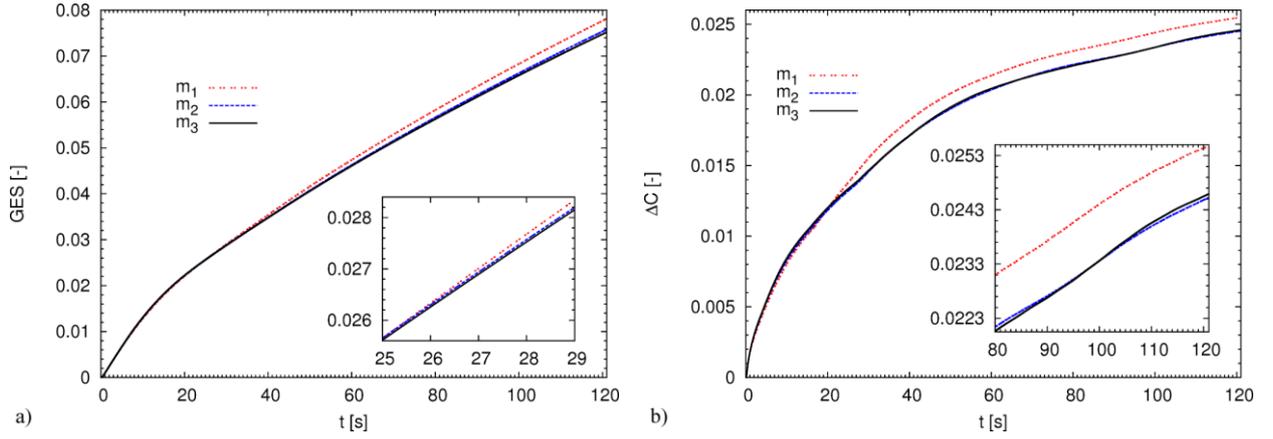

Figure 9. Histories of the convergence of a) the global extent of segregation GES and b) maximal concentration differences $\Delta C$ on three gradually refined grids $m_1: 64 \times 256$, $m_2: 196 \times 384$, $m_3: 128 \times 512$ CV's. The convergence was obtained and mesh independent solution is localized close to the solution obtained on grid $m_3$.

As is shown in Figure 8bd, a similar dependence is predicted for $C_M$ profiles. In this case, the convergence is achieved in the solid ($x < 2.2\ cm$) and in the mushy zone ($2.2\ cm \leq x \leq 6.1\ cm$) regions at both times depicted in Figure 8. As in the case of $g_S$ profiles at $t = 100\ s$ the convergence of the $C_M$ in pure fluid ($x > 6.1\ cm$) is more evident than at the final simulation time $t = 121\ s$ (compare Figure 8b-d). This behavior is caused by the presence of the right wall and, as mentioned above, different rates of advancement of the liquidus solidification fronts. The convergence of the solid fraction $g_S$ and the mixture concentration $C_M$ depicted in Figure 8 confirms the convergence of temperature and velocity fields (see Eq. (1) and Eq. (7) respectively). In Figure 9, histories of the convergence of the GES and $\Delta C$ macrosegregation monitoring functions are presented. We notice that in both cases the convergence on gradually refined grids is achieved. GES can be interpreted as a second order norm and $\Delta C$ as a first order norm, they are measures of the distance between solutions on three gradually refined grids. As is shown in Figure 9b, the convergence of the solute distribution measured by $\Delta C$ is less evident, particularly at the initial stages of solidification when $t \leq 24\ s$. However, the solute distributions obtained on the finest mesh $m_3$ can be considered as sufficiently close to the mesh independent solution.

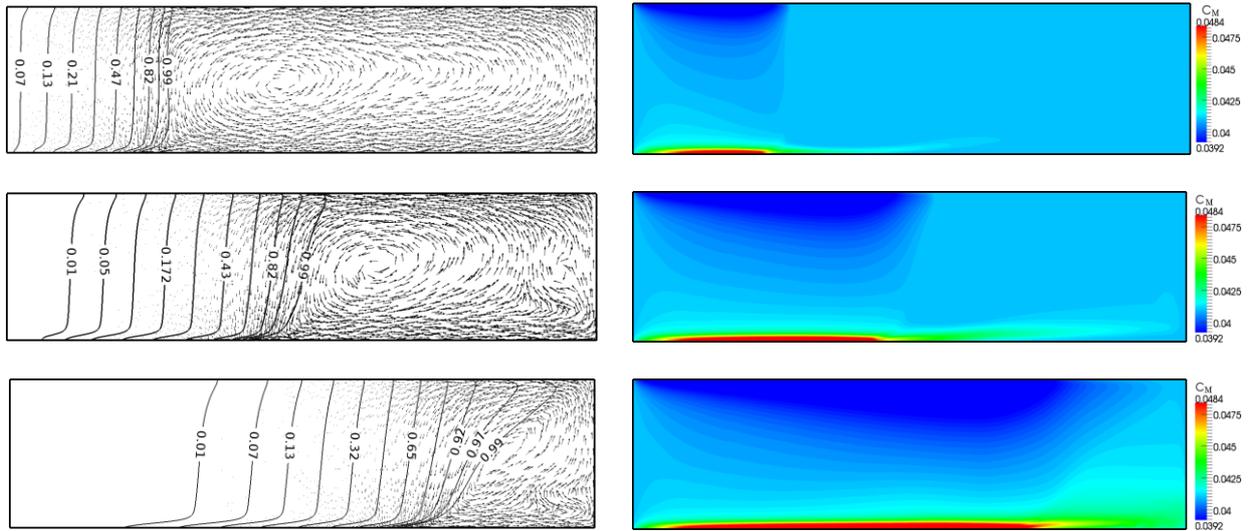

Figure 10. Velocity fields and $g_L$ iso-lines (left), $C_M$ distributions (right) at three time moments $t = 31s, 66s, 121s$ from top to bottom. The concentration of the copper (solute) is enriched close to the bottom wall and depleted close to the top wall of the cavity.

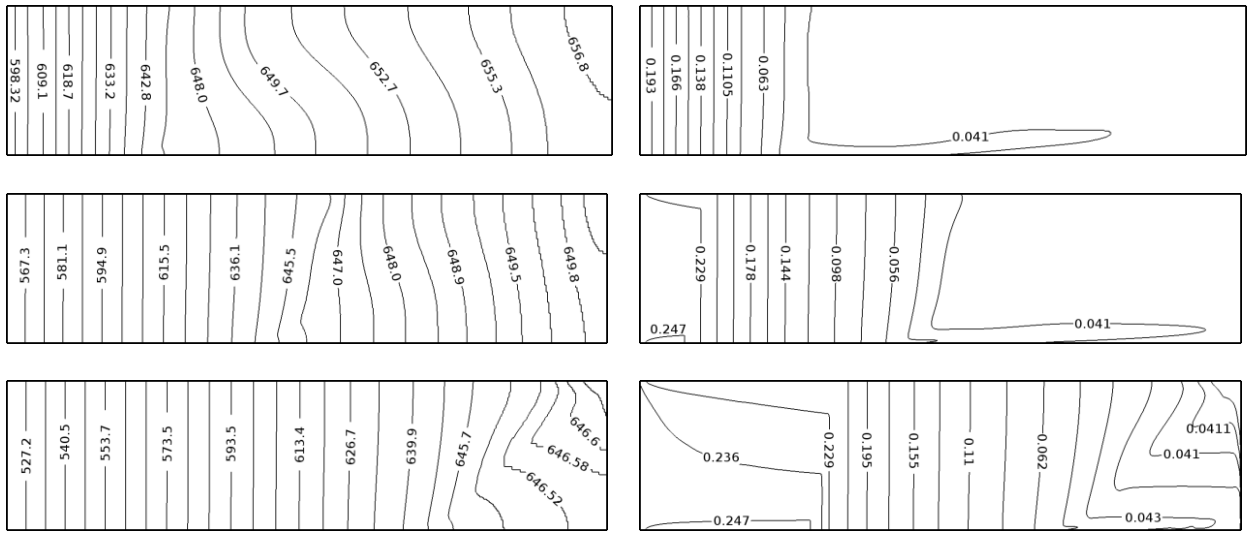

Figure 11. The isotherms (left) and $C_L$ iso-lines (right) at time moments $t = 31s, 66s, 121s$ from top to bottom, initially $C_L = C_I = 0.041$.

In Figures 10-11, a solute jet inducing macrosegregation and influencing the shape of solidification fronts can be observed. Since the solute rejected during solidification of aluminum is transported and deposited at the bottom wall of the cavity, the velocity field induced by the counter clock-wise rotating vertex washes away deposited solute and forms a jet. The solute jet is responsible for the modification of the solidification fronts velocities (compare Figure 10 and Figure 11). This leads to an increment of the solute concentration in a bulk fluid and a significant deceleration of the solidification process at the bottom wall of the cavity. At the same time, depletion of the solute concentration close to the top wall enhances the advancement of the solidification fronts.

**Conclusions**

In this paper, a verification study of the binary-fluid solidification model coupled with the finite volume flow solver Star-CD is presented. In the first part, we derive the novel semi-analytical solution to the heat diffusion equation. Afterwards, the range of applicability of the novel semi-analytical solution is analyzed demonstrating that it can be used for verification of the numerical model only when $St > 100$; for smaller Stefan numbers the non-linear effects related to the latent heat rejection become dominant. Hence, the linearization introduced to integrate the heat diffusion equation is not sufficient approximation of the binary-fluid solidification process for large latent heat magnitudes. In the second part, the macrosegregation model coupled with the flow solver, is verified, showing convergence of the main variables on the gradually refined grids. The correct coupling between binary-fluid solidifcation model and the flow solver guarantees reliable prediction of the solidification fronts positions as well as the macrosegregation patterns.

**Appendix A**

In this appendix, we describe the change of variables in Eq. (11) inside the mushy zone. The non-dimensional temperature $\theta$ in the solid, the mushy zone and the liquid is given by Eqs. (11). After substitution in to Eq. (11) we obtain

$$\left(\frac{1}{\alpha_M} - \frac{\rho L_{ht}}{k_M}\frac{\partial g_S}{\partial T}\right)\frac{\partial \theta_M}{\partial t} = \frac{\partial^2 \theta_M}{\partial x^2} + \frac{1}{k_M}\frac{\partial k_M}{\partial x}\frac{\partial \theta_M}{\partial x}. \tag{A1}$$

The similarity variable $\eta = xg(t)$, $g(t) = 1/(2\sqrt{\alpha_S t})$ is used to rewrite temporal and spatial derivatives in Eq. (A1) in the non-dimensional form. The time derivative in (A1) becomes

$$\frac{\partial \theta_M}{\partial t} = \frac{\partial \theta_M}{\partial \eta}\frac{\partial \eta}{\partial t} = -2\alpha_S \eta g^2 \frac{\partial \theta_M}{\partial \eta}. \tag{A2}$$

The first and second order spatial derivatives read

$$\frac{\partial \theta_M}{\partial x} = \frac{\partial \theta_M}{\partial \eta}\frac{\partial \eta}{\partial x} = \frac{\partial \theta_M}{\partial \eta}g, \tag{A3}$$

$$\frac{\partial}{\partial x}\left(\frac{\partial \theta_M}{\partial x}\right) = \frac{\partial}{\partial \eta}\left(\frac{\partial \theta_M}{\partial \eta}\frac{\partial \eta}{\partial x}\right)\frac{\partial \eta}{\partial x} = \frac{\partial^2 \theta_M}{\partial \eta^2}g^2. \tag{A4}$$

The second term on the RHS in Eq. (A1) can be written as

$$\frac{1}{k_M}\frac{\partial k_M}{\partial x}\frac{\partial \theta_M}{\partial x} = \frac{1}{k_M}\frac{\partial k_M}{\partial \eta}\frac{\partial \theta_M}{\partial \eta}g^2. \tag{A5}$$

Using Eqs. (A1-A5) and notation introduced in Eq. (16) we obtain Eq. (15).

**Appendix B**

Here, we derive the Stefan conditions at the solidus front $x = x_S$, inside the mushy zone at $x = x_l$, $l = 1, \ldots, N - 1$ and at the liquidus front $x = x_L$. Let us integrate Eq. (9) in the $N + 1$ infinitesimal control volumes $\delta x_k = x_k^+ - x_k^- \to 0$ where $k = S, l, L$, respectively. Since the jumps of temperature $[T] = 0$ at $x_S, x_l, x_L$ and the jumps of the solid fraction $[g_S] = 0$ at $x_l$ and $[g_S] \neq 0$ at $x_S, x_L$; integrated in $N + 1$ control volumes Eq. (9) reads

$$at\ x = x_S\ :\ -\rho L_{ht}\frac{dx_S}{dt}\int_{x_S^-}^{x_S^+}\frac{\partial g_S}{\partial x}dx = \int_{x_S^-}^{x_S^+}\frac{\partial}{\partial x}\left(k\frac{\partial T}{\partial x}\right)dx \tag{B1}$$

$$at\ x = x_l\ :\ \int_{x_l^-}^{x_l^+}\frac{\partial}{\partial x}\left(k\frac{\partial T}{\partial x}\right)dx = 0 \tag{B2}$$

$$at\ x = x_L\ :\ -\rho L_{ht}\frac{dx_L}{dt}\int_{x_L^-}^{x_L^+}\frac{\partial g_S}{\partial x}dx = \int_{x_L^-}^{x_L^+}\frac{\partial}{\partial x}\left(k\frac{\partial T}{\partial x}\right)dx \tag{B3}$$

where we make an assumption that $dx_k/dt = const.$ in $\delta x_k \to 0$ for $k = S, L$. After spatial integration, taking into account the extraction of heat flux from the semi-infinite domain at $x = 0$ and orientation of normal vectors at the faces of the 1D control volumes $\delta x_k$, we obtain

$$at\ x = x_S\ :\ \rho L_{ht}\frac{dx_S}{dt}(1 - g_{S,sol}) = k_S\frac{\partial T_S}{\partial x} - k_M^l\frac{\partial T_M^l}{\partial x} \tag{B4}$$

$$at\ x = x_l\ :\ \frac{\partial T_M^l}{\partial x} - \frac{\partial T_M^{l+1}}{\partial x} = 0 \tag{B5}$$

$$at\ x = x_L\ :\ \rho L_{ht}\frac{dx_L}{dt}g_{S,liq} = k_M^{l+1}\frac{\partial T_M^{l+1}}{\partial x} - k_L\frac{\partial T_L}{\partial x} \tag{B6}$$

where $g_{S,sol} = g_S(T_{sol})$, $g_{S,liq} = g_S(T_{liq})$ and $l = 1, \ldots, N - 1$. Next, we show how non-dimensional temperatures $\theta$ and similarity variable $\eta$ are introduced into Eqs. (B4-B6). Let us first note that the velocities of the solidification fronts may be rewritten as

$$at\ x = x_S\ :\ \frac{dx_S}{dt} = \frac{1}{g}\frac{\partial \eta_S}{\partial t} - \frac{\eta_S}{g^2}\frac{\partial g}{\partial t} = 2\alpha_S g\eta_S \tag{B7}$$

$$at\ x = x_L\ :\quad \frac{dx_L}{dt} = \frac{1}{g}\frac{\partial \eta_L}{\partial t} - \frac{\eta_L}{g^2}\frac{\partial g}{\partial t} = 2\alpha_S g \eta_L \tag{B8}$$

since we are searching for a stationary solution to Eq. (11) at large times $t$. The non-dimensional temperatures in the solid, the mushy zone $k-th$ interval and in the liquid read

$$\theta_S = \frac{T - T_C}{T_{sol} - T_C},\qquad \theta_M^k = \frac{T - T_{k-1}}{T_F - T_{k-1}},\qquad \theta_L = \frac{T - T_I}{T_{liq} - T_I}. \tag{B9}$$

where $k = 1, \ldots, N$ and where for $k = 1$: $T_0 = T_{sol}$. Substitution of Eqs. (B7-B9) in Eqs. (B4-B6) results in

$$\theta_E \frac{\partial \theta_S}{\partial \eta} - r_{MS}^1 \frac{\partial \theta_M^1}{\partial \eta} = \frac{2\eta_S(1 - g_{S,sol})}{St_0} \tag{B10}$$

$$\theta_{sol}^l \frac{\partial \theta_M^l}{\partial \eta} - \frac{\partial \theta_M^{l+1}}{\partial \eta} = 0 \tag{B11}$$

$$r_{MS}^N \frac{\partial \theta_M^N}{\partial \eta} - \theta_I r_{LS} \frac{\partial \theta_L}{\partial \eta} = \frac{2\eta_L g_{S,liq}}{St_{N-1}} \tag{B12}$$

where $l = 1, \ldots, N-1$, $r_{MS}^l = k_M^l/k_S$, $St_0 = c_S(T_F - T_{sol})/L_{ht}$, $\theta_{sol}^l = (T_F - T_{l-1})/(T_F - T_l)$ where for $l + 1 = N$: $T_N = T_{liq}$, $\theta_I = (T_{liq} - T_I)/(T_F - T_{N-1})$, $St_{N-1} = c_S(T_F - T_{N-1})/L_{ht}$.

**Appendix C**

In this appendix, we show how to determine the equivalent temperatures $T_{eq}^k$ inside of the $k = 1, \ldots, N$ intervals introduced in the mushy zone for averaging. The temperature averaged solid fraction gradient is calculated in the $k - th$ temperature interval

$$\langle \frac{\partial g_S}{\partial T} \rangle \Big|_k = \frac{1}{\Delta T_k} \int_{T_{k-1}}^{T_k} \frac{\partial g_S}{\partial T} dT,\quad \Delta T_k = T_k - T_{k-1} \tag{C1}$$

where for $k = 1$: $T_0 = T_{sol}$ and for $k = N$: $T_N = T_{liq}$. Since the temperature gradient of the solid fraction $g_S$ defined in Eq. (1) reads

$$\frac{\partial g_S}{\partial T} = \frac{1}{1 - k_P}\left[\frac{1}{T - T_F} - \frac{T - T_{liq}}{(T - T_F)^2}\right], \tag{C2}$$

Eq. (C1) after integration becomes

$$\langle \frac{\partial g_S}{\partial T} \rangle \Big|_k = \frac{1}{(1 - k_P)\Delta T_k}\left(\frac{T_k - T_{liq}}{T_k - T_F} - \frac{T_{k-1} - T_{liq}}{T_{k-1} - T_F}\right),\quad \Delta T_k = T_k - T_{k-1}. \tag{C3}$$

One notices that since all temperatures in Eq. (C3) are known, the expression given by Eq. (C3) is constant in each interval $k$. Therefore, using the definition given by Eq. (17) and Eq. (C2), we can write

$$A_k = (1 - k_P) \langle \frac{\partial g_S}{\partial T} \rangle \bigg|_k = \left[ \frac{1}{T_{eq}^k - T_F} - \frac{T_{eq}^k - T_{liq}}{(T_{eq}^k - T_F)^2} \right]. \tag{C4}$$

Finally, from Eq. (C4) the equivalent temperature inside the $k-th$ temperature interval reads

$$T_{eq}^k = T_f - \sqrt{\frac{T_{liq} - T_F}{A_k}}. \tag{C5}$$

One notices that when $k = N$ and $T_N = T_{liq}$ then $A_N = 1/(T_{N-1} - T_F)$ and

$$T_{eq}^N = T_F - \sqrt{(T_{liq} - T_F)(T_{N-1} - T_F)}, \tag{C6}$$

in particular for $N = 1$ we obtain Eq. (18) from [5].

**Appendix D**

Here, we describe the solution procedure for the set of $N + 1$ non-linear equations obtained from the conditions for heat fluxes in the mushy zone. This solution is required to find the unknowns $\eta_S, \dots, T_l, \dots, \eta_L$ where $l = 1, \dots N - 1$ denotes a number of unknown temperatures $T_l$ at points

$$\eta_l = \eta_S + \frac{l}{N}(\eta_L - \eta_S) \tag{D1}$$

and $N \geq 2$ is the number of temperature intervals introduced for averaging. For simplicity, all following equations are derived using the lever rule model given by Eq. (1). First, we note the introduction of $N$ averaging intervals $\Delta T_l = T_l - T_{sol}, \dots, \Delta T_k = T_k - T_{k-1}, \dots, \Delta T_N = T_{liq} - T_{N-1}$ where for k= 1 : $T_0 = T_{sol}$ and for $k = N$ : $T_N = T_{liq}$ increases the number of the heat diffusion equations that must be solved to obtain piecewise temperature profiles. The set of $N + 2$ ODE's that must be solved reads

$$2\eta \frac{\partial \theta_S}{\partial \eta} + \frac{\partial^2 \theta_S}{\partial \eta^2} = 0, \qquad 0 \leq \eta \leq \eta_S \tag{D2}$$

$$(2\eta a_k + b_k) \frac{\partial \theta_M^k}{\partial \eta} + \frac{\partial^2 \theta_M^k}{\partial \eta^2} = 0, \qquad \eta_{k-1} \leq \eta \leq \eta_k \tag{D3}$$

$$2\eta \frac{\alpha_S}{\alpha_L} \frac{\partial \theta_L}{\partial \eta} + \frac{\partial^2 \theta_L}{\partial \eta^2} = 0, \qquad \eta_L \leq \eta \leq \infty \tag{D4}$$

where for $k = 1$ : $\eta_0 = \eta_S$ and for $k = N$ : $\eta_N = \eta_L$. Analytical solutions to Eqs. (D2-D4) are given by nondimensional temperature profiles, written employing the notation introduced in Eq. (40)

$$\theta_S = \frac{erf(\eta)}{erf(\eta_S)} \quad : \quad 0 \leq \eta \leq \eta_S \tag{D5}$$

$$\theta_M^k = \theta_{liq}^k \frac{erf[d_k(\eta)] - erf[d_k(\eta_{k-1})]}{erf[d_k(\eta_k)] - erf[d_k(\eta_{k-1})]} \quad : \quad \eta_S \leq \eta \leq \eta_L \tag{D6}$$

$$\theta_L = \frac{erfc(\sqrt{b}\eta)}{erfc(\sqrt{b}\eta_L)} \quad : \quad \eta_L \leq \eta \leq \infty \tag{D7}$$

where $\theta_M^k, \theta_{liq}^k$ read

$$\theta_M^k = \frac{T - T_{k-1}}{T_F - T_{k-1}}, \qquad \theta_{liq}^k = \frac{T_k - T_{k-1}}{T_F - T_{k-1}}. \tag{D8}$$

The above formulation assures the continuity of temperature at points $\eta_S, \ldots, \eta_l, \ldots, \eta_L$, where $l = 1, \ldots, N-1$ and $k = 1, \ldots, N$.

In the second step, we determine constants $a_k, b_k$ present in Eqs. (D6), see also Eq. (37). Knowing $T_{eq}^k$ from Eq. (C5) for $k = 1, \ldots, N$ allows the calculation of the representative solid fractions and material properties inside each of the k-th temperature intervals

$$g_S^k = \frac{1}{1 - k_P} \frac{T_{eq}^k - T_{liq}}{T_{eq}^k - T_F} \tag{D9}$$

$$c_M^k = g_S^k c_S + (1 - g_S^k) c_L \tag{D10}$$

$$k_M^k = g_S^k k_S + (1 - g_S^k) k_L \tag{D11}$$

and the further coefficients $\alpha_S/\alpha'_S$ in Eq. (15) can be also determined

$$a_k = \left.\frac{\alpha_S}{\alpha'_S}\right|_k = \frac{c_M^k - L_{ht} \left.\langle\frac{\partial g_S}{\partial T}\rangle\right|_k}{c_S \frac{k_M^k}{k_S}}. \tag{D12}$$

The contribution from the heat conductivity coefficient gradient $b_k$ is approximated by central differences for $k = 2, \ldots, N-1$

$$b_k = \left.\frac{1}{k_M}\frac{\partial k_M}{\partial \eta}\right|_k \approx \frac{1}{k_M^k}\frac{k_M^{k+1} - k_M^{k-1}}{\eta_M^{k+1} - \eta_M^{k-1}} = \frac{k_S - k_L}{k_{M,eq}^k}\frac{g_{S,eq}^{k+1} - g_{S,eq}^{k-1}}{\eta_M^{k+1} - \eta_M^{k-1}} \tag{D13}$$

or forward and backward differences for $k = 1$ or $k = N$ respectively, see Eqs. (34-35). The unknown positions $\eta_M^k$ are determined as a linear combination of $\eta_S$ and $\eta_L$

$$\eta_M^k = \eta_{k-1} + \frac{1}{2}(\eta_k - \eta_{k-1}) \tag{D14}$$

where $\eta_k$ is obtained from Eq. (D1).

In the third step we formulate the set of non-linear equations required to determine the unknowns $\eta_S, \ldots, T_l, \ldots, \eta_L$. After substitution of Eqs. (D5-D7) into the boundary conditions for heat fluxes given by Eqs. (B10-B12); the set of $N + 1$ non-linear equations read

$$\frac{\theta_E exp(-\eta_S^2)}{erf(\eta_S)} - \frac{r_{MS}^1 \theta_{liq}^l \sqrt{a_1} exp(-d_1(\eta_S)^2)}{erf(d_1(\eta_1)) - erf(d_1(\eta_S))} = \frac{\sqrt{\pi}\eta_S(1 - g_{S,sol})}{St_0} \tag{D15}$$

$$\frac{\theta_{sol}^l \theta_{liq}^l \sqrt{a_l} exp(-d_l(\eta_l)^2)}{erf(d_l(\eta_l)) - erf(d_l(\eta_{l-1}))} = \frac{\theta_{liq}^{l+1} \sqrt{a_{l+1}} exp(-d_{l+1}(\eta_l)^2)}{erf(d_{l+1}(\eta_{l+1})) - erf(d_{l+1}(\eta_l))} \tag{D16}$$

$$\frac{r_{MS}^N \theta_{liq}^N \sqrt{a_N} exp(-d_N(\eta_L)^2)}{erf(d_N(\eta_L)) - erf(d_N(\eta_{N-1}))} + \frac{\theta_l r_{LS} \sqrt{b} exp(-b\eta_L^2)}{erfc(\sqrt{b}\eta_L)} = \frac{\sqrt{\pi}\eta_L g_{S,liq}}{St_{N-1}} \tag{D17}$$

where $g_{S,sol}$, $g_{S,liq}$ are solid volume fractions obtained for the temperatures $T_{sol}$ and $T_{liq}$, see Eq. (1). The solution of the set of Eqs. (D15-D17) using non-linear equation solver allows to obtain the vector of $N + 1$ unknowns $\eta_S, \ldots, T_l, \ldots, \eta_L$ where $l = 1, \ldots, N - 1$. After substitution of $\eta_S, \ldots, T_l, \ldots, \eta_L$ into the non-dimensional temperatures profiles given by Eqs. (D5-D7) the final solution is obtained.

## Acknowledgements

This work was funded by the German Research Foundation (DFG) in the framework of the project "Multi-phase-based modeling, simulation and experimental validation of mold filling and solidification of metallic melts in the light of foreign particles and porosity." SCHA: 814/132, AOBJ: 577070. We are also grateful to Prof. M. Peric and Dr. Huang Jianbo from CD-Adapco for discussions and technical support.

## Nomenclature

$c$ - specific heat, $J \cdot kg^{-1} \cdot K^{-1}$

$f$ - mass fraction

$g$ - volume fraction

$g_i$ - i-th gravitational acceleration component, $m \cdot s^{-2}$

$h$ - mixture enthalpy, $J \cdot kg^{-1}$

$k$ - thermal conductivity, $W \cdot m^{-2} \cdot K^{-1}$

$k_P$ - partition coefficient

$m$ - total mass, $kg$

$m_L$ - slope of liquidus line, $K \cdot \%^{-1}$

$p$ - pressure, $N \cdot m^{-2}$

$q$ - heat flux, $W \cdot m^{-2}$

$t$ - time moment, $s$

$u_i$ - i-th velocity component, $m \cdot s^{-1}$

$x$ - spatial coordinate, $m$

$x_S$ - position of the solidus line, $m$

$x_L$ - position of the liquidus line, $m$

$A$ - auxiliary constant

$B$ - solvent concentration, $m$

$C$ - solute concentration, $\%$

$D$ - diffusion coefficient, $m^2 \cdot s^{-1}$

$K$ - permeability coefficient, $m^2$

$K_0$ - relative permeability coefficient, $m^2$

$L$ - horizontal dimension of the computational domain, $m$

$L_{ht}$ - latent heat of fusion, $J \cdot kg^{-1}$

$N$ - number of temperature intervals

$N_c$ - number of CV's

$T$ - temperature, $K$

$V$ - volume of representative control volume, $m^3$

$St$ - Stefan number

**Greek symbols**

$\alpha$ - thermal diffusivity, $m^2 \cdot s^{-1}$

$\eta$ - similarity variable

$\theta$ - non-dimensional temperature

$\lambda$ - secondary dendrite arm spacing, $m^2$

$\mu_L$ - liquid alloy dynamic viscosity, $kg \cdot m^{-1} \cdot s^{-1}$

$\rho$ - mixture density, $kg \cdot m^{-3}$

$\chi$ - thermal expansion coefficient

$\psi$ - solutal expansion coefficient

$\Delta C$ - difference between the maximum and minimum solute concentration, %

**Subscripts/Superscripts**

$C$ - cold wall

$E$ - eutectic

$F$ - fusion

$I$ - initial

$L$ - liquid

$M$ - mixture

$S$ - solid

$eq$ - equivalent

$liq$ - liquidus

$sol$ - solidus

**Abbreviations**

CFL - Courant-Friedrich-Levy condition

GES - global extent of segregation

ODE - ordinary differential equation

**References**

Bars L.M., Worster M.G., Interfacial conditions between pure fluid and a porous medium, implications for binary alloy solidification, *J. Fluid Mech.*, 550, pp. 149-173 2006,

Bennon W.D., Incropera F.P., A continuum model for momentum, heat and species transport in binary solid-liquid phase change systems--I. Model formulation, *Int. J. of Heat Mass Transf.*, vol. 30, pp 2161—2170, 1987

Bennon W.D., Incropera F.P., A continuum model for momentum, heat and species transport in binary solid-liquid phase change systems--II. Application to solidification in rectangular cavity, *Int. J. of Heat Mass Transfer*, vol. 30, pp. 2161—2170, 1987

Braga S.L., Viskanta R., Solidification of a binary solution on a cold isothermal surface, Int. J. Heat Mass Transfer 33 (4) (1990) 745-754

Chakaraborty S., Dutta P., An analytical solution for conduction dominated unidirectional solidification of binary mixtures. *Appl. Math. Modeling*, vol. 26, pp. 545—561, 2002

Jakumeit J., Jana S., Wacławczyk T., Mehdizadeh A., Youani J., and Buehrig-Polaczek A., Four-phase fully coupled mold-fling and solidification simulation for gas porosity prediction in aluminum sand casting. In 13th MCWASP, Austeria, June 17-22, 2012.

Kurz W., Fischer D.J., *Fundamentals of Solidification*, Trans Tech Publications (1990)

Samanta, S., Zabaras, N., Numerical study of macrosegregation in aluminium alloys solidifying on uneven surfaces, *Int. J. of Heat Mass Transfer*, vol. 85, pp. 481—501, 2004

Schäfer M., *Computational Engineering, Introduction to Numerical Methods*, Springer, Berlin-Heidelberg-New York, 2006

Voller V.R., Brent A.D., The modeling of heat, mass and solute transport in solidification systems, *Int. J. of Heat Mass Transfer*, vol. 32, pp. 1719—1731, 1989

Voller V.R., A similarity solution for the solidification of a multicomponent alloy, Int. J. Heat Mass Transfer 40 (12) (1997) 2869-2877

Worster M.G., Solidification of an alloy from a cooled boundary, *J. Fluid Mech.*, vol. 167, pp. 481—501, 1986

Wang Chao-Yang, Beckermann C., A two-phase mixture model of liquid-gas flow and heat transfer in capillary porous media—I. Formulation, *International Journal of Heat and Mass Transfer,* Volume 36, Issue 11, July 1993, Pages 2747-2758

Wang Chao-Yang, Beckermann C., A two-phase mixture model of liquid-gas flow and heat transfer in capillary porous media—II. Application to pressure-driven boiling flow adjacent to a vertical heated plate, *International Journal of Heat and Mass Transfer,* Volume 36, Issue 11, July 1993, Pages 2759-2768

Wacławczyk T., Sternel D., and Schäfer M.. Verification of binary fluid solidification model. In ICNAAM AIP Conference Proc., volume 1389, Halkidiki, Greece, 2011.